# Correlated excited states in the narrow gap band semiconductor FeSi and antiferromagnetic screening of local spin moments.

Sergii Khmelevskyi[1], Georg Kresse[2], and Peter Mohn[1].

[1]Center for Computational Materials Science, Institute for Applied Physics, Vienna University of Technology, Wiedner Hauptstrasse 8, A-1040, Vienna, Austria.

[2]Computational Materials Physics, University of Vienna, Sensengasse 8/12, A-1090 Vienna, Austria.

The physical properties of the semiconductor FeSi with very narrow band gap, anomalous behavior of the magnetic susceptibility and metal-insulator transition at elevated temperatures attract gross interest due to the still controversial theoretical understanding of their origin. On one side the purely band like mechanism of the gap formation in FeSi at low temperature is well established, on other side a number of experiments and their theoretical interpretation suggest a rich physics of strong correlations at finite temperature. In this work we use an ab-initio scheme based on the Random Phase Approximation and Local Spin Density Approximation (RPA@LSDA) to reveal the role of the electron correlation effects in FeSi extending it by applying a fixed spin moment constraint. In the parameter free framework we show that correlation effects essentially alter the one-electron LSDA results leading to the formation of an additional state with finite magnetic moment on Fe, whose energy is almost degenerate with the non-magnetic ground state. This explains the results of high field experiments, which found a first-order meta-magnetic phase transition into a metallic ferromagnetic state. Our results suggest a strongly correlated nature of the low-energy excitations in FeSi. From our super-cells calculations we reveal that these excitations are local and exhibit a Kondo-like behavior since a strong antiferromagnetic screening is present.

# I. Introduction.

The study of the metal-insulator transition in the narrow gap semiconductor FeSi has been one of the most intriguing stories in the field of itinerant magnetism, the physics of electron correlations in solids, and the band theory of the solid state for more than four decades. From the very beginning it has been understood[1] that the insulating state in FeSi with B20 crystal structure is due to a gap within the Fe d-states caused by the hybridization with the Si p-states. Band structure calculations,[2] including numerous modern first-principles investigations, readily confirm this observation predicting a narrow band gap of about 100-150 meV within the mean-field like Local Density Approximation[3,4,5,6] (LDA) and the Hartree-Fock approach.[7] Thus, for a long time, till the 90ies, FeSi has been considered as an ordinary band insulator with a very narrow gap where the gap formation occurs due to one-electron effects rather than being a Mott insulator,[8] where strong electron correlation is the reason for the insulating state. This has put FeSi out of the scope of the physics of electron correlations for decades. However, a number of quite distinctive features have attracted considerable experimental and theoretical attention even during these times. First-of all, there is the unusual temperature behavior of the magnetic susceptibility, which exhibits a pronounced maximum at about 500 K, followed by a Curie-Weiss like decay at high temperature suggesting the existence of local magnetic moments, and an almost constant value at low temperatures until 80-100 K where a fast increase is observed.[9] In their seminal work Jaccarino et al. [9] have shown that two different models may account for this experimental susceptibility behavior: i) electron excitations between two extremely narrow bands, below and above of the Fermi level, and ii) local excitations between Fe singlet (S=0) and spin-doublet (S=1/2) or spin-triplet (S=1) states. The simple two narrow band model has been further shaped by the results of modern band-structure calculations to provide a good fit to the experimental results on specific heat and susceptibility.[10] Jacarrino's second model of local excitations or "spin-transitions" has never been confirmed and remained abandoned in the further extensive discussion of the properties of FeSi during the last decades (see Ref. 11). However, despite the absence of a theoretical justification for the origin of the local doublet or triplet states and due to difficulties in the interpretation of model parameters, the local model still provides the best fit9 to the experimental specific heat and high temperature susceptibility. In the present work based on parameter free total energy calculations in an ab-initio framework, we provide evidence for the existence of an excited many-body state with spin-½ that is separated from the ground state non magnetic singlet by a small energy difference.

The itinerant electron spin-fluctuation theory based on the model density of states has been used by Moriya and Takahashi[12] to account for high temperature susceptibility behavior. In the framework of the spin-fluctuation theory FeSi is a nearly ferro-, or antiferro- magnetic semiconductor and its susceptibility maximum and Curie-Weiss like high temperature behavior is due to thermally induced fluctuating Fe moments. Since then, FeSi has become an almost iconic system in spin-fluctuation theory of itinerant electron magnetism[13,14] providing the background for testing its various approximations for a long time.[15,16,17,18] The success of spin-fluctuation theory in explaining the high temperature susceptibility behavior, inelastic neutron scattering results, and the prediction of the more or less correct band gap value and peaky DOS structure around the Fermi level from band structure calculations have led to the conclusion[5,14] that band theory provides an adequate description of the physical properties of FeSi, except at elevated temperature where dynamical correlation effects are important for the adequate description of the metal-insulator transitions and temperature delay of the susceptibility growth.[14,19] Indeed modern first-principles based studies of the electronic structure that account for correlation effects within dynamical mean-field theory (LDA+DMFT)[20,21,22] predict simultaneously a closing of the band gap with increasing temperature and describe the temperature of the susceptibility maximum. However, these calculations also reveal a few problems. While with a reasonable choice of the correlation parameters the position of the susceptibility maximum is correct, the value of the susceptibility appears to be a few times smaller than in experiment.[21] Although one can increase the correlation parameter (in DMFT) to obtain the right order of magnitude of the susceptibility, at the same time the agreement between theoretical and experimental susceptibility curves becomes worse[22] (see also model calculations in Ref. 23). The correlation effects were found to be also important to explain the experiments on the thermal evolution of the optical conductivity.[4,24] Optical spectroscopy experiments together with low temperature resistivity measurements and their interpretation pointed in the direction of FeSi as a strongly correlated insulator.

## II. Dichotomy of FeSi problem: "heavy fermions" versus "band insulator".

It was noted[25] that the activated low temperature behavior of the resistivity and susceptibility, which quantitatively cannot simply be accounted for by the Fermi distribution function, is quite similar to the phenomenology of correlated Kondo insulators. Infrared spectroscopy measurements[26,27] reveal that for FeSi some for an ordinary band semiconductor quite unexpected features exist, such as low frequency spectral weight loss and unusual temperature dependence of the gap appearance which can indeed be interpreted as a "Kondo insulator" behaviour.[28] This suggestion has triggered quite intense experimental and theoretical efforts to understand the low temperature properties of FeSi. The experiments and theoretical arguments pointing towards a "Kondo insulator" model have been summarized in the review article by L. Degiorgi.[29] In particular, it was argued[30] that there is clear evidence for Anderson localization in the optical spectra of FeSi and an overall similarity to condensed heavy Fermion systems. It must be noted though that Angular Resolved Photoemission Spectroscopy (ARPES) studies do not prove the Kondo insulator ground state of FeSi.[31,32,33] However, these experiments point towards an essential role of the electron correlation for the metal insulator transition at finite temperature. The state-of-the art theoretical investigations based on LDA+DMFT during the last decades also conclude that the correlation effects within the Fe d-bands lead to an essential renormalization of the electronic structure and are the driving mechanism for the metal-insulator transition at finite temperature.[20,21,22,34,35] However, they also support the conclusion of the mean-field like band structure theory that the insulating ground state of FeSi is due to ordinary hybridization effects and thus FeSi is a band insulator. Among others, this dichotomy of the FeSi results gave rise to a novel field of research named "correlated band insulator"[20] or "narrow gap semiconducting problem".[29] Indeed the canonical experimental work by S. Paschen et al.[36] on the low temperature properties of FeSi performed on high quality samples has revealed a plethora of anomalies in the low-temperature behavior of the susceptibility, resistivity, magneto-resistance and specific heat below the metallization temperature. The full account of these properties cannot be derived from a simple one-electron picture, even assuming a temperature depended gap value, and only very few of them can be explained by the effects of impurities and Fe structural defects.[36] It is interesting to note that e.g. the Schottky-type anomalies in the low temperature susceptibility and specific heat, may point towards the existence of localized spin-state transitions and an Anderson localization picture. The temperature dependence of the effective conducting gap, as well as its difference in value from the optical gap imposes a particular problem. Degiorgi et al.[30] argued that the sharp narrow absorption peaks in the optical spectra of FeSi might be due to a transition between "mid-gap" localized states and the continuum of the conduction and valence bands, thus advocating the picture of localized excitonic states. Later the experimental and theoretical arguments have pointed[37] towards the formation of spin-polaronic states in FeSi. Recent THz radiation probe measurements revealed a broad absorption peak which appears below 20K. This has been interpreted as a formation of spin-polaronic states in the middle of the gap due to strong correlation effects.[38] Thus, there is a direct evidence of some many-body states formed at low temperature in the mid-gap region of the FeSi electronic structure. In this work we will argue that these energetically localized many-body states might be also localized in real space providing the possibility for spin-state transitions and the emergence of Kondo-like physics.

FeSi and doped $Fe_{1-x}TM_xSi$ alloys (TM – transition metals) have also attracted a huge interest due to their unusual thermoelectric properties.[35,39,40] In particular, recently, this research lead to the observation of phonon softening and an anomalous downshift of the acoustic peak with temperature.[41,42,43] This has been ascribed to the metallization, which supports the idea[5,44] that the metal-insulator transition in FeSi might be driven by the effect of thermal lattice disorder.[42] Very recent state-of-the-art first-principles calculations of the phonon spectra and the thermally averaged electronic structure concluded that the gap in FeSi indeed closes at elevated temperatures due to increasing thermal disorder[45] thus providing a conceptually different scenario for the metal-insulator transition (see also the discussion in Ref.46). One can thus summarize that, despite of the great successes of spin fluctuation theory, LDA calculations and modern methods of correlated electron theories (LDA+DMFT), the problem of the metal-insulator transition and the excitation spectra of FeSi is far from being completely understood. Nowadays, the great importance of longitudinal spin-fluctuations at elevated temperatures above the metal-insulator transition and electron-electron correlation effects at low temperatures as well as the pure band character of the insulating state at zero temperature are commonly agreed. However, the large dissimilarity of the two band-gap values - obtained from spectroscopic and transport measurements - the "heavy fermion" like behavior, and the relevance of the "Kondo" physics at elevated temperatures, and the exact mechanism driving the metal-insulator transition, and the character of the collective excitations seen at low temperatures in the "mid-gap" region, are still debated.

## III. Methodology and Motivation.

In this work we approach the problem in the following way. First, we investigate the correlated electronic structure of FeSi using a parameter free ab-initio approach based on the GW approximation[47,48] and calculate the total energy dependence on the local Fe magnetic moment within the Random Phase Approximation (RPA)[49] and compare it to the LDA result. We find that the RPA produces a separated high-spin state (with a spin moment of exactly one Bohr magneton per Fe) whose energy is almost degenerate, but slightly lower than the energy of the non-magnetic state. The energy difference (~0.14 meV/Fe) is few orders of magnitude smaller than the PBE calculated energy of the magnetic moment formation in typical band ferromagnetic metals, like bcc Fe ( 565 meV/Fe) and fcc Ni ( 46 meV/Ni). This result is quite similar to the LDA+U results derived by Anisimov *et al.*[10] by tuning the empirical U-parameter. Our parameter free calculations thus provide a first-principles evidence that in FeSi there exist magnetic states with the energies very close to the non-magnetic ground state. In a second step, we assume that such correlated excited state might be localized in space, and explore the consequences of this assumption by performing large super-cell calculations where some of the Fe atoms are constrained to be in the excited correlated high-spin state, whereas the rest of the system is allowed to converge freely. We find a notable result: up to a certain concentration of the excited atoms, the total magnetic moment of the super-cell is zero since the high-spin state of the "excited" Fe becomes fully screened by the neighboring Fe atoms exhibiting anti-parallel spin polarization. Such a situation is a precursor for a Kondo-singlet formation in the excited state. We argue that the situation in FeSi might be the following: the ground state is a band-insulator singlet, whereas intrinsic elementary excitations at low temperatures are screened Kondo singlets. This observation might reconcile the "heavy fermion" vs. band insulator dichotomy of the FeSi problem.

In the present work all calculations were done using the Vienna Ab-initio Simulation Package (VASP)[50] which is based on the Projector Augmented Wave method[51] in the implementation of Kresse and Joubert.[52] On the Density Functional Theory level we employ the Perdew, Burke, Ernzerhof (PBE)[53] exchange-correlation potential to obtain the electronic band structure and a Kohn-Sham orbital basis for the GW calculations and the evaluation of the total energy in the RPA. The GW quasi-particle spectrum has been obtained in a way described by Shishkin and Kresse.[54] The evaluation of the RPA correlation energy rests on the application of the adiabatic connection fluctuation-dissipation theorem.[55,56] The RPA energy is added to the total Hartree-Fock (HF) energy, one electron energy + exact Hartree-Fock exchange (EXX), calculated on the PBE converged Kohn-Sham orbitals to obtain the total energy of the system. This procedure is described in detail in Ref.[57] and due to the use of PBE orbitals it is called EXX+RPA@PBE approximation. This approach has recently proven to be an effective tool to study the impact of electronic correlations in solids.[58,59,60,61,62] From a technical point of view, up to our best knowledge, the specialty of our application of the EXX+RPA@PBE approximation here is its combination with the fixed spin moment (FSM) constraint.[63] Apart from the consequences it has for our present study of FeSi, we believe that it provides the most straightforward way to generalize the application of EXX+RPA@PBE (or the ...@PBE/LSDA family of methods) for studies of magnetic solids.

## IV. Band Theory of the Ground State and Magnetism.

In Fig. 1 we show the self-consistent PBE Density of States (DOS) of FeSi for the cubic B20 structure calculated for the experimental lattice parameters.[36] Since this DOS is very well known,[3,5,45] we show only the most interesting region around the gap at the Fermi level. The value of the PBE gap is 130 meV, which is in good agreement with previous LDA based calculations,[3,5,24] and only somewhat slightly larger than the experimentally estimated values (~50-110 meV).[36] Basically the agreement between the PBE gap values is almost perfect considering the optical spectroscopic estimates, which are close to 100 meV, rather than the values derived from fits to the resistivity or susceptibility which give 50-60 meV. This agreement has been always regarded as surprising since it is well known that PBE potential grossly underestimate the gap value for almost all band semiconductors.[19] Reasonable one-particle gap values are usually obtained within the GW method, which for band semiconductors gives quite accurate quasiparticle spectra and increases the PBE gap.[64] In the case of FeSi the calculated GW gap (~220 meV) is almost two times larger than the PBE one (see Fig. 1) making the agreement between the theoretical and experimental gap worse. However, the PBE prediction of the narrow gap in FeSi, has always been a strong argument for the "band" insulator rather than a "Kondo" insulator origin of the ground state. PBE predicts very narrow DOS peaks just below and just above the gap edges. The latter nicely support spectroscopic observations as well as the widely used two-narrow-band model. In addition it shows that the narrow peak in the valence band near the Fermi level is of purely one-electron ("non-Kondo") origin. The GW DOS indeed preserves all these DOS features leading to a just moderate DOS renormalization compared to PBE. This result is basically in line with conclusions drawn from LDA+DMFT studies[20,21,34] describing FeSi as a moderately correlated electron system.

Although LDA and PBE correctly predicts the insulating non-magnetic ground state of FeSi the finite temperature behavior of the susceptibility points towards a non-trivial magnetism in this material, which has been explained on different levels of model descriptions assuming the existence of temperature induced magnetic moments.[12,14,34] Since FeSi is regarded as a nearly ferro- or antiferromagnetic material the dependence of the (free) energy on the local Fe magnetic moment amplitude must play a central role in any itinerant electron theory of the magnetic properties of this material, e.g. via a Stoner exchange enhancement factor, an on-site Hubbard – U, or being just straightforwardly calculated from first-principles. Moreover, for nearly ferromagnetic materials with a susceptibility maximum at elevated temperatures, the itinerant electron theory predicts a meta-magnetic transition to a magnetically ordered state for a sufficiently high applied field.[65] Indeed for FeSi the dependence of the total energy on the value of the Fe moments has been calculated from Local Spin Density Approximation (LSDA) several times and a meta-magnetic transition has been predicted.[10,66] However, as has been shown by Yamada *et al*.[67], that the predicted critical field magnitude depends on the size of the LSDA gap and moreover itinerant electron magnetism theory applied using the calculated LSDA energy vs. magnetic moment predicts two meta-magnetic transitions – the first at a few hundred Teslas (T) to a weak ferromagnetic state (~0.1 $\mu_B$/Fe) and the second at about[67] 700 T or[10] 1000 T to the high moment state (~1 $\mu_B$/Fe). The FSM total energy E(m) dependence calculated in the PBE approximation is shown in Fig. 2. The energy increases monotonously up to a value of 1 $\mu_B$ /Fe where it has an inflection point and increases its slope considerably.

Since PBE is known to favor magnetism as compared to the LSDA used in [67] our PBE FSM energies (Fig. 2) give smaller values for the meta-magnetic critical fields about 200 T for transition into the weak ferromagnetic state (with ~0.1 $\mu_B$/Fe) and 350 T for the transition into the high spin state with a moment of about ~0.8 $\mu_B$/Fe. The values of the critical field was calculated numerically using data presented in the fig.2 by adding the magnetic energy in the external field. Experimentally the ultra-high field experiment gives only one meta-magnetic transition in fields at about 355±20 T, at 4.2 K, and the transition is into the metallic high spin state with a Fe moment of ~0.95 $\mu_B$.[68] Although PBE predicts the transition into the high-spin state with reasonable accuracy, the absence of the transition into the weak ferromagnetic state points to the importance of correlations and quantum effects. The experimental moment in the high spin state is close to a spin-1/2 state (1 $\mu_B$/Fe), however, PBE predicts only 0.8 $\mu_B$/Fe. The latter is also suspicious owing to the fact that PBE tends to overestimate the moment, in particular, considering that the high-spin state is metallic both in theory and experiment.

The finding by Anisimov et al.[10] was that the energy vs. moment curve for FeSi changes considerably if one alters LSDA by adding some local U corrections in the framework of the LSDA+U approach.[69] By increasing the phenomenological U parameter they found a second minimum in the E(m) curve at m= 1 $\mu_B$/Fe and the ferromagnetic metallic state becomes the ground state for a rather high value of U ~4.6 eV. Basically it is not surprising that increasing the local U-parameter one can drive a system into a magnetic state. However, the interesting fact is that in FeSi increasing of the U-parameter drives the system into a metallic state as well, which is completely opposite to the Mott metal-insulator scenario. Since an *a priory* choice of the proper U-parameter is hardly possible, the authors[10] give an interpretation of their results in terms of a two-band correlation model.

## V. Consequences of the RPA correlations.

From the previous discussion there emerges enough motivation to explore the role of the correlation effects on the magnetism of FeSi using a parameter free ab-initio scheme. To this end we calculate the energy vs. moment curve for FeSi using the EXX+RPA@PBE method. On the model level (two-band Hubbard model with model DOS) the importance of the RPA for understanding of the low temperature properties of the FeSi has been mentioned by Takahashi.[14] To apply RPA on the first-principles basis, we obtain a self-consistent PBE electronic structure for different values of the Fe magnetic moments and then calculate the exact exchange energy and the RPA correlation energy as described in Ref.[59]. The resulting energy vs. moment curve is presented in Fig. 2. One can immediately see that the behavior of the EXX+RPA@PBE total energy is *qualitatively* different from PBE. There is a second minimum exactly at a moment value of 1 $\mu_B$/Fe with an energy just 14 meV/Fe below the non-magnetic ground state. The most important result of the RPA treatment of correlations is that it eliminates the possibility of an intermediate transition into a weakly ferromagnetic state, suggesting that the system will undergo the transition immediately into the high spin state. The position of the high-spin RPA minimum is in good agreement with experiment[68] giving 1 $\mu_B$/Fe and predicting the metallic character. The very existence of the two isolated energy minima provides strong arguments toward a two-states transition model, which as mentioned before, much better describes the specific heat data than the two-band model.[9] These arguments will be further refined and justified in the next section. However, the first-principle EXX+RPA@PBE predicts that a high-spin state is lower in energy than non-polarized state, but for a very small energy difference. It is known that the RPA somewhat overestimates the energy difference between of the spin-polarized and non-spin-polarized solutions and some corrections beyond the RPA scheme, suggested recently, could shift the high-spin energy in metallic state upwards.[70,71,72] For instance, as shown recently by Maggio and Kresse[73] the random phase approximation overestimates the correlation energy of the 3D electron gas because of the neglect of second order screened exchange. However, these corrections are too expensive to be calculated at present for realistic solid material.

In Fig. 3 we separately plot the two contributions to the the EXX+RPA@PBE energy (from Fig.2) - HF energy and the RPA correlation energy. One can see two distinctive regions where exchange and correlations affects magnetism in a different way This is evident from the behavior of the two curves in Fig.3 – at lower moments RPA correlations favor magnetic state with respect of non-magnetic (m=0) whereas HF energy favors magnetic ground state at higher moments. At low moments, where the system is insulating, the correlation energy favor magnetism, but the one-electron (HF) part makes the system non-magnetic. At higher moments (in the metallic state) the high moment state becomes stabilized due to HF exchange interactions whereas correlations tend to reduce the magnetization. Thus one can conjecture that in the metallic state FeSi is a correlated Hund's metal, whereas PBE fails to predict the existence of the high-moment state. On the other hand, the non-magnetic insulating ground state of FeSi is due to the one-electron HF contribution that makes the FeSi a band insulator since the non-exchange part of total HF energy is dominating.

#
### VI. Structure of the local excitations in FeSi.

In this section, our study will be based on the observation that the results similar to the EXX+RPA@PBE behavior of the E(m) curve can be also obtained in LSDA+U calculations for some values of the U-parameter. The authors of Ref.10 correctly note that the metallic high spin-state considered in their LDA+U calculation is not compatible with a local ionic description. Since the same argument also holds for our RPA result, we are going to investigate the consequence of the local magnetic excitations, essentially of many-body character, on the Fe sites on the electronic structure of FeSi. To this end we assume that part of the Fe atoms may be excited into a high-spin state with 1 $\mu_B$/Fe, whereas the rest are in the initial uncorrelated non-magnetic ground state, but their electronic state might be affected by the excited iron atoms. Our assumption has the following reasoning: i) the shape of the EXX+RPA@PBE E(m) curve (Fig. 2) suggests that, at least at low temperatures, only two states might be physically relevant namely those with S=0 and S=1/2; ii) optical spectroscopy experiments (see above), suggest the existence of energetically localized mid-gap states in FeSi at low temperatures; iii) the spin-transition model provides the best fit to the experimental specific heat measurements. The main idea of our theoretical construction for the study of the lowest energy excitations in FeSi is essentially similar to that used in the well known work on the γ-α transition in fcc Cerium, where the mixed valence state has been modeled by an alloy of Ce atoms with a correlated and uncorrelated f-shell.[74,75] Our goal, however, is more limited since we want to explore how the electronic structure of FeSi would react on the local atomic excitations. Moreover, the utility of our model will become justified *a posteriori* by the transparency of the obtained results.

We construct a 2x2x2 cubic super cell containing 32 Fe and 32 Si atoms. For a few randomly chosen (but out of the 1$^{st}$ neighbor shell of each other) Fe atoms in the super-cell, we apply the PBE+U approximation,[76] with U = 2.2 eV for the d-electrons. U=2.2 eV is the value that produces an *E(m)* curve close to the RPA result (see Fig. 3 where the corresponding calculated PBE+U results are shown) and such that the high-spin solution is slightly lower in energy. The rest of the system is treated within the conventional PBE functional. The calculations converge to a spin-polarized solution for all chosen super-cells. The "excited" Fe atoms, where the PBE+U potential has been applied, possess moments around 1.6 $\mu_B$ in calculations done for supercells with one, two, three or four such "excited" atoms per 32 total Fe atoms. However, the Fe atoms surrounding the "excited" Fe have a negative spin polarization (~0.1-0.3 $\mu_B$) whereas more distant Fe neighbours have small positive or nearly zero polarizations, such that the total magnetic moment of the system remains zero. The calculated iso-surfaces of positive and negative spin polarizations are plotted in the Figs. 4a and 4b for supercells with one and three "excited" Fe atoms, correspondingly. It appears that the large magnetic moments on the "excited" Fe atom cause a spin density wave on the "normal" Fe atoms leading to a full antiferromagnetic screening. Such a situation is a precursor for the Kondo effect in diluted metallic magnetic alloys. In the case of FeSi it might appear that the antiferromagnetic screening of local magnetic excitations would lead to the formation of many-body Kondo singlet states. The excitations of such many-body singlet states at low temperature indeed might explain the band-insulator-"Kondo" or "heavy-fermions" dichotomy of the FeSi behavior at low temperatures. One may speculate that the peculiarity of the FeSi band semiconductor is a strongly correlated character of the excitonic states due to the correlated d-states participating in its formation. Indeed one can view the states in figures 4a and 4b as localized Frenkel-type excitons.[77] By increasing the concentration of excited Fe atoms to 5 per supercell, the picture of the correlated localized excitons breaks down, and the calculations converge to the ferromagnetic state as shown in Fig. 4c. In this case, the regions with negative spin-polarization are almost absent and the large spin-moments of the excited atoms are no longer screened. In Fig.5 we schematically show the "spin-polarization wave" around the central "excited" Fe atom. It is interesting to see that the scenario of local excitations is also able to describe the metal-insulator transition. In Fig. 6 we plot the density of states calculated for five supercells with 1-5 "excited" atoms. One can see that "excitons" tend to reduce the PBE gap (~130 meV – Fig.1). By exceeding a certain concentration (5 out 32 Fe atoms excited), they lead to metallization and also to the vanishing of the full antiferromagnetic screening and thus the condition for the formation of Kondo states. It is interesting to note that Co substitution in B20 $Fe_{1-x}Co_xSi$ alloys, which leads to a metallic and ferromagnetic ground state,[78] induces a positive spin-polarization on the neighboring Fe atoms, as has been shown in the calculations of Mazurenko *et al.*[21] for a 64 atom supercell with one Co impurity. Thus a Co impurity is unable to form a Kondo-like singlet state unlike the locally excited Fe atoms.

Let us also note that even starting from antiferromagnetic configurations of the excited Fe our calculations converged to the "ferromagnetic" configurations with total zero moment as shown in Fig. 4. This might suggest a ferromagnetic coupling between excited centers, but any final conclusion concerning the possible cooperative effects between localized "excitons" cannot be drawn at present.

**VII. Summary**

On the basis of parameter free simulations, we have shown that the correlation effects on the RPA level lead to a stable metallic high-spin state in FeSi, whose energy is very close to the non-magnetic ground state with a narrow band gap. The inclusion of the explicit correlation effects in parameter free manner leads to the appearance of two isolated energy minima – a non-spin polarized insulating state and a high-spin ($1\mu_B$ per iron) metallic state. Assuming that correlations might also lead to local atomic excitations in the Fe-d shells we show that such excitations will be antiferromagnetically screened by the neighboring Fe atoms, which is a precursor for a possible many-body Kondo singlet formation and experimentally observed "heavy fermion"-like physics of FeSi at low temperatures. Our theoretical construction should be considered as a call for a more refined theoretical model to describe the excited states in FeSi, which might be complimentary to the modern strongly correlated theories based on the Hubbard model. It is most important to make it clear, that a realistic model for finite temperature effects in FeSi should provide the dynamical treatment of the excitations in semiconducting FeSi beyond the static picture used in this paper. However, and this is the main conclusion which can be drawn from the study presented in section 6, this dynamical picture should go beyond the single-site approximations used so far in theories of the correlation effects in FeSi.

The application of the first principle based RPA scheme to FeSi has shown that the dichotomy of the FeSi properties originates from the existence of two energetically competing states, each exhibiting a different role of the correlation effects. We argue that the structure and the origin of the low temperature "mid-gap" states in FeSi also might have a correlated character and need to be further verified by experiment. Our study demonstrates, in particular, that application of the fixed spin moment constraint in EXX+RPA@PBE family of methods might be crucial for the investigation of magnetism in correlated solids.

The work has been supported by the Centre of Computational Materials Science (SK) and the Austrian Science Fund (FWF): F4109-N28 SFB ViCoM (PM, GK).

**Figure 1. The ground DOS of FeSi around Fermi level calculated with PBE (black) and GW (red) approximation (colour online).**

**Figure 2.** The total energy dependence on the magnetic moment in FeSi calculated with PBE (black) and EXX+RPA@PBE (red) approximations. The energy is given per Fe atoms (colour online). The stars shows the energies calculated in PBE+U method with U=2.2 eV.

**Figure 3.** The contributions to the EXX+RPA@PBE total energy (per Fe) in B20 FeSi: black Hartree-Fock contribution, red – RPA correlation energy (colour online).

**Figure 4.** The iso-surfaces of the spin-polarizations for 64 atoms cell of FeSi as described in the text. The yellow surface is positive spin-polarization, blue is negative. The brown spheres are Fe atoms, the dark blue spheres are Si atoms. The Fe atoms surrounded by large positive iso-surface are "excited" (see text) Fe atoms. a) one "excited" Fe in the supercell, b) three "excited" Fe atoms per supercell and c) five Fe "excited" atoms per supercell (colour online).

**Figure 5.** The "schematic representation" of the spin-polarization around the central Fe exited atoms (red) in 64 atomic supercell with one exited atom. White and Yellow circles representing the Fe atoms in the first and second NN shell of the exited Fe site. The numbers in the figure is the average atomic spin in the shell.

**Figure 6.** Calculated DOS around the Fermi level for a 64 atom supercell with one to five "excited" Fe atoms (colour online).

+

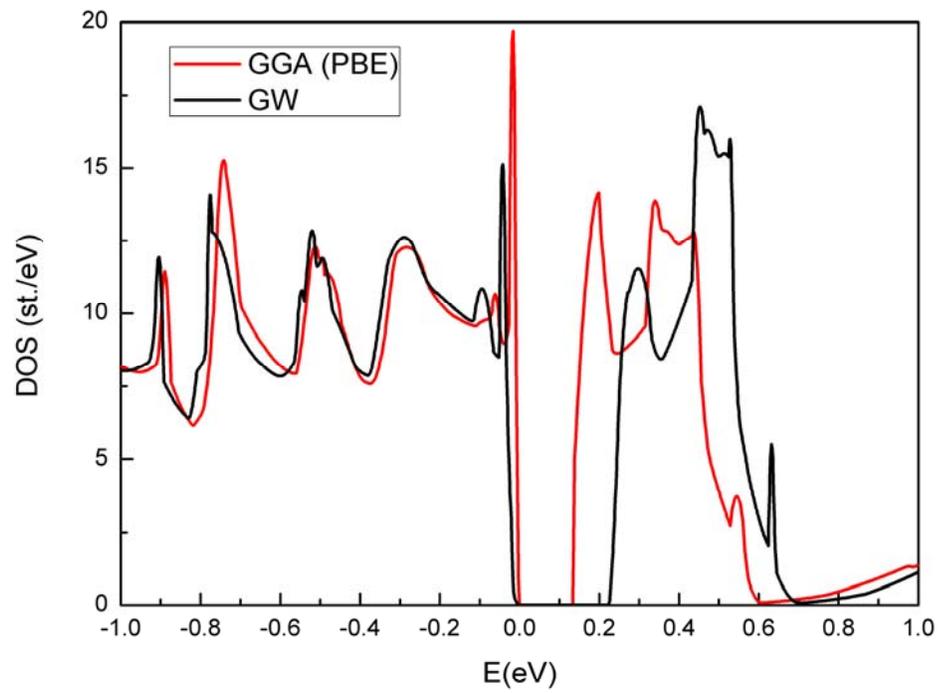

FIGURE 1_KHMELEVSKYI

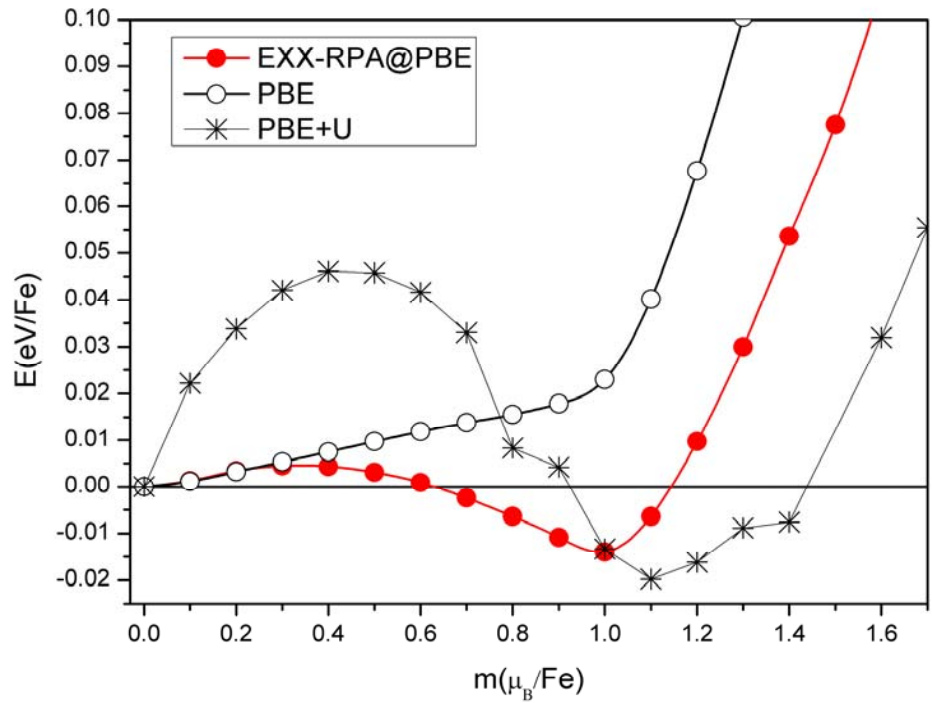

FIGUER2_KHMELEVSKYI

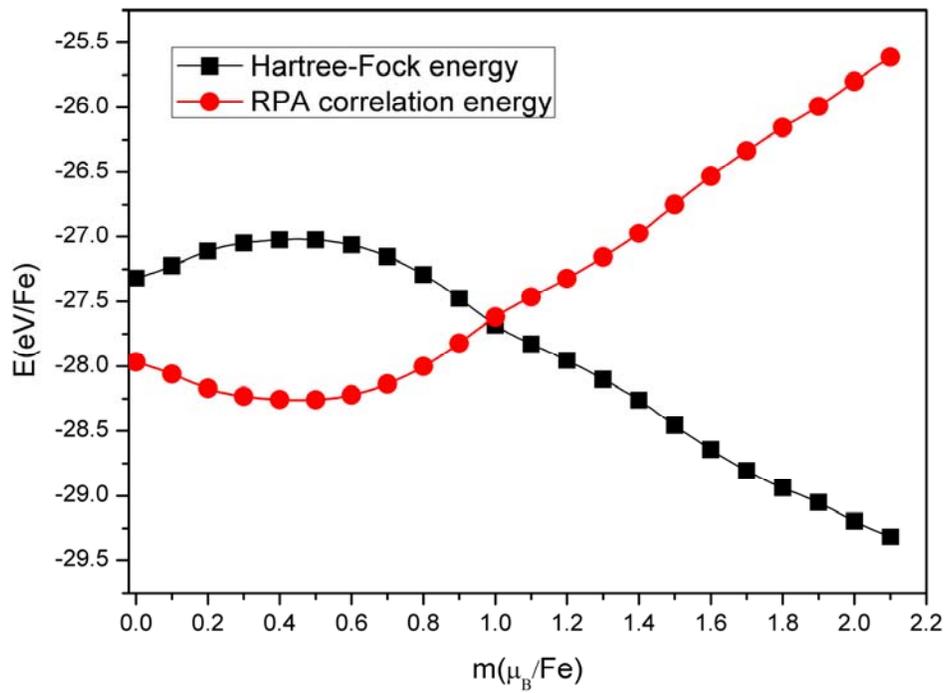

FIGURE 3_KHMELEVSKYI

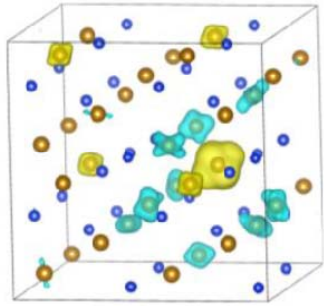

a)

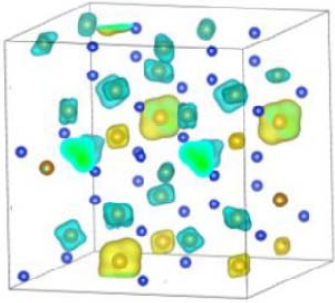

b)

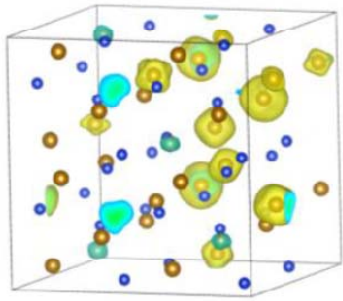

c)

FIGURE 4_KHMELEVSKYI

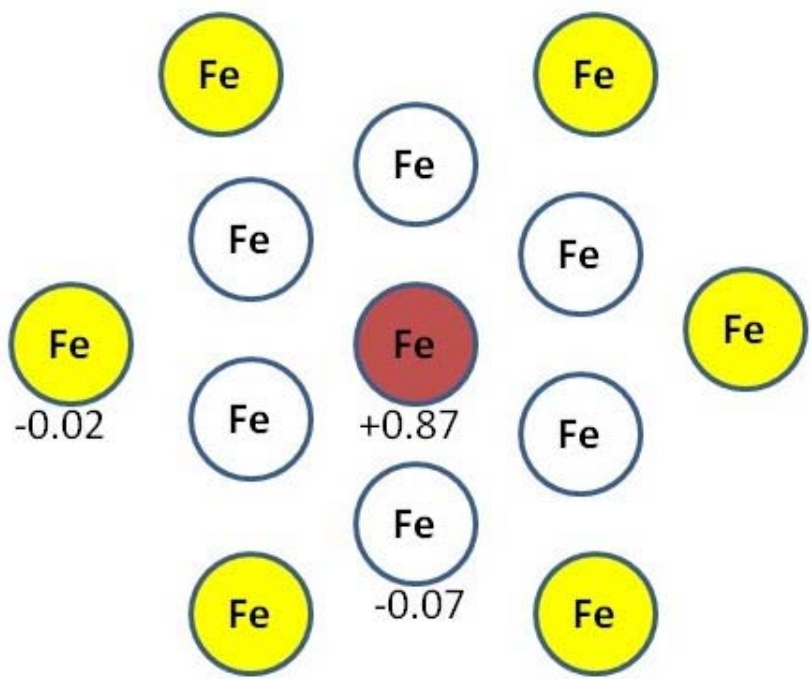

FIGURE 5_KHMELEVSKYI

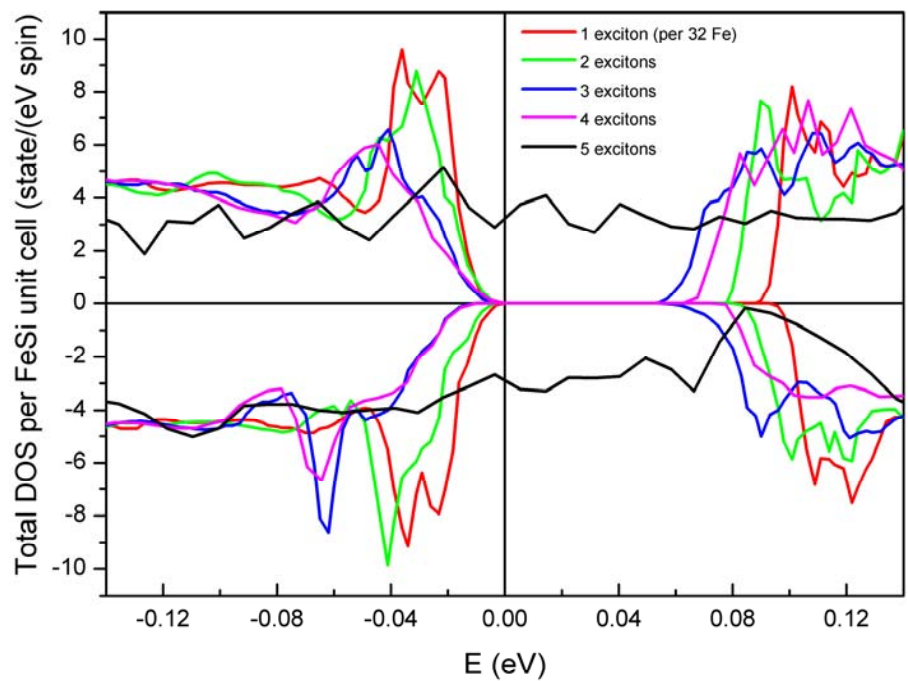

FIGURE 6_KHMELEVSKYI

**References.**

[1] L. Pauling, A.M. Soldate, Acta Cryst. **1**, 212 (1948).

[2] O. Nakanishi, A. Yanase, **A.** and A. Hasegawa J. Magn. Magn. Mater. **15-18**, 879 (1980).

[3] L. F. Mattheiss, and D. R. Hamann, Phys. Rev. B **47,** 13114 (1993)

[4] C. Fu, M. P. C. M. Krijn, and S. Doniach, Phys. Rev. B **49,** 2219 (1994).

[5] T. Jarlborg, Phys. Rev. B **59,** 15002 (1999).

[6] E. G. Moroni, W. Wolf, J. Hafner, and R. Podloucky, Phys. Rev. B **51,** 1286 (1999).

[7] M. Neef, K. Doll and G. Zwicknagl, J. Phys.: Condens. Matter **18,** 7437–7447 (2006).

[8] N. F. Mott, *Metal-Insulator transitions* (Taylor&Francis, London, New York, Philadelphia, 1990).

[9] V. Jaccarino, G. K. Wertheim, J. H. Wernick, L. R. Walker, and S. Arajs, Phys. Rev. **160**, 476 (1967).

[10] V. I. Anisimov, S. Yu. Ezhov, I. S. Elfimov, I. V. Solovyev, and T. M. Rice, Phys. Rev. Lett. **76**, 1735 (1996).

[11] D. van der Marel, A. Damascelli, K. Schulte, A. A. Menovsky, Physica B **244**, 138 (1998)

[12] T. Moriya, and Y. Takahashi, J. Phys. Soc. Japan **45** 397 (1978).

[13] T. Moriya, *Spin Fluctuations in Itinerant Electron Magnetism* (Berlin: Springer, 1985)

[14] Y. Takahashi, J. Phys.: Condens. Matter **9,** 2593 (1997).

[15] S. N. Evangelou and D. M. Edwards, J. Phys. C: Solid State Phys. **16,** 2121 (1983).

[16] H. Yamada, H. Oka, Physica B **337,** 170 (2003).

[17] A. A. Povzner, A. G. Volkov, and P. V. Bayankin, Phys. of Sol. Stat. **40**, 1305 (1998)

[18] A. A. Povzner, A. G. Volkov, T. A. Nogovitsyna, J. of Magn. and Magn. Mater. **409**, 1 (2016).

[19] T. Jarlborg, Phys. Rev. B **51**, 11106 (1995).

[20] J. Kuneš, and V. I. Anisimov, Phys. Rev. B **78**, 033109 (2008).

[21] V. V. Mazurenko, A. O. Shorikov, A. V. Lukoyanov, K. Kharlov, E. Gorelov, A. I. Lichtenstein, and V. I. Anisimov, Phys. Rev. B **81**, 125131 (2010).

[22] Y. Yanagi, and K. Ueda, Phys. Rev. B **93**, 045125 (2016).

[23] K. Urasaki and T. Saso, Phys. Rev. B **58**, 15 528 (1998).

[24] C. Fu, and S. Doniach, Phys. Rev. B **51**, 17439 (1995).

[25] G. Aeppli, and Z. Fisk, Commnets Condens. Matter Phys. 15, 155 (1992).

[26] Z. Schliesenger, Z. Fisk, H. T. Zhang, M. B. Maple, J. F. Tusa, G. Aeppli, Phys. Rev. Lett **71**, 1748 (1993).

[27] H. Ohta, S. Kimura, E. Kulatov, S. Halilov, T. Nanba, M. Motokawa, M. Sato, K. Nagasaka, J. Phys. Soc. Japan **83**, 4206 (1994).

[28] Z. Schliesenger, Z. Fisk, H. T. Zhang, M. B. Maple, Physica B **237-238**, 460 (1997).

[29] L. Degiorgi, Rev. Mod. Phys. 71, 687 (1999).

[30] L. Degiorgi, M. B. Hunt, H. R. Ott, M. Dressel, B. J. Feenstra, G. Grüner, Z. Fisk and P. Canfield, Europhys. Lett **28**, 341 (1994).

[31] M. Klein, D. Zur, D. Menzel, J. Schoenes, K. Doll, J. Röder, and F. Reinert, Phys. Rev. Lett **101**, 046406 (2008).

[32] M. Arita, K. Shimada, Y. Takeda, M. Nakatake, H. Namatame, M. Taniguchi, H. Negishi, T. Oguchi, T. Saitoh, A. Fujimori, and T. Kanomata, Phys, Rev. B **77**, 205117 (2008).

[33] D. Zur, D. Menzel, I. Jursic, and J. Schoenes, L. Patthey, M. Neef, K. Doll, and G. Zwicknagl, Phys. Rev. B **75**, 165103 (2007).

[34] J. M. Tomczak, K. Haule, and G. Kotliar, Proc. Natl. Acad. Sci. USA **109**, 3243 (2012).

[35] J. M. Tomczak, K. Haule, and G. Kotliar, ch. **4**, pp. 45-57, in *New Materials for Thermoelectric Applications: Theory and Experiment* (ed. by V. Zlatic, and A. Hewson, Springer, Dordrecht, 2013).

[36] S. Paschen, E. Felder, M. A. Chernikov, L. Degiorgi, H. Schwer, H. R. Ott, D. P. Young, J. L. Sarrao, and Z. Fisk, Phys. Rev. B **56**, 12916 (1996).

[37] N. E. Sluchanko, V. V. Glushkov, S. V. Demishev, M. V. Kondrin, K. M. Petukhov, N. A. Samarin, V. V. Moshchalkov and A. A. Menovsky, Europhys. Lett. **51**, 557 (2000).

[38] V. V. Glushkov, B. P. Gorshunov, E. S. Zhukova, S. V. Demishev, A. A. Pronin, N. E. Sluchanko, S. Kaiser, and M. Dressel, Phys. Rev. B **84**, 073108 (2011).

[39] B. C. Sales, O. Delaire, M. A. McGuire, and A. F. May, Phys. Rev. B **83**, 125209 (2011).

[40] B. C. Sales, E. C. Jones, B. C. Chakoumakos, J. A. Fernandez-Barca, H. E. Harmon, and J. W. Sharp, Phys. Rev. B **50**, 8207 (1994).

[41] O. Delaire, K. Marty, M. B. Stone, P. R. C. Kent, M. S. Lucas, D. L. Abernathy, D. Mandrus, and B. C. Sales, Proc. Natl. Acad. Sci. USA **108**, 4725 (2011).

[42] O. Delaire, I. I. Al-Qasir, J. Ma, A. M. dos Santos, B. C. Sales, L. Mauger, M. B. Stone, D. L. Abernathy, Y. Xiao, and M. Somayazulu, Phys. Rev. B **87**, 184304 (2013).

[43] S. Krannich, Y. Sidis, D. Lamago, R. Heid, J.-M. Mignot, H.v. Loehneysen, A. Ivanov, P. Steffens, T. Keller, L. Wang, E.Goering, and F. Weber, Nat. Commun. **6**, 8961 (2015).

[44] T. Jarlborg, Phys. Rev. B **76**, 205105 (2007).

[45] R. Stern, and G. K. H. Madsen, Phys. Rev. B **94**, 144304 (2016)

[46] P. P. Parshin, A. I. Chumakov, P. A. Alekseev, K. S. Nemkovski,, J. Perßon, L. Dubrovinsky, A. Kantor, and R. Rüffer, Phys. Rev. B **93**, 081102(R) (2016).

[47] L. Hedin, Phys. Rev. **139**, A796 (1965).

[48] F. Aryasetiawan and O. Gunnarsson, Rep. Prog. Phys. **61**, 237 (1998).

[49] F. Furche, Phys. Rev. B **64**, 195120 (2001).

[50] G. Kresse and J. Furthmüller, Comput. Mater. Sci. **6**, 15 (1996).

[51] P. E. Blöchl, Phys. Rev. B **50**, 17953 (1994)

[52] G. Kresse and D. Joubert, Phys. Rev. B **59**, 1758 (1999).

[53] J. P. Perdew, K. Burke, and M. Ernzerhof, Phys. Rev. Lett. **77**, 3865 (1996).

[54] M. Shishkin and G. Kresse, Phys. Rev. B **74**, 035101 (2006).

[55] O. Gunnarsson and B. I. Lundqvist, Phys. Rev. B **13**, 4274 (1976).

[56] D. C. Langreth and J. P. Perdew, Solid State Commun. **17**, 1425 (1975).

[57] J. Harl and G. Kresse, Phys. Rev. B **77**, 045136 (2008).

[58] J. Harl and G. Kresse, Phys. Rev. Lett. **103**, 056401 (2009)

[59] J. Harl, L. Schimka, and G. Kresse, Phys. Rev. B **81**, 115126 (2010).

[60] L. Schimka, J. Harl, A. Stroppa, A. Grüneis, M. Marsman, F. Mittendorfer, and G. Kresse, Nature Materials **9**, 741 (2010).

[61] E. Trushin, M. Betzunger, S. Blügel, and A. Göring, Phys. Rev. B **94**, 075123 (2016).

[62] B. Ramberger, T. Schäfer, and G. Kresse, Phys. Rev. Lett. **118**, 106403 (2017).

[63] L. M. Sandratskii. Adv. Phys. **47**, 91 (1998).

[64] M. Shishkin, M. Marsman, and G. Kresse, Phys. Rev. Lett. **99**, 246403 (2007).

[65] P. Mohn and S. Khmelevskyi. in *Band-Ferromagnetism: Ground-State and Finite Temperature Phenomena*, edited by K. Baberschke, M. Donath, W. Nolting, p.126-142 (Springer, Berlin, New York, Tokyo, 2001)

[66] E. Kulatov, H. Ohta, J. Phys. Soc. Japan **66**, 2386 (1997) 2386.

[67] H. Yamada, K. Terao, H. Ohta, T. Arioka and E. Kulatov, Physics B **281-282**, 267 (2000).

[68] Yu. B. Kudasov A. I. Bykov, M. I. Dolotenko, N. P. Kolokol'chikov, M. P. Monakhov, I. M. Markevtsev, V. V. Platonov, V. D. Selemir, O. M. Tatsenko, A. V. Filippov, A. G. Volkov, A. A. Povzner, P. V. Bayankin, V. G. Guk, and V. V. Kryuk, JETP **89**, 960 (1999).

[69] V. I. Anisimov, F. Aryasetiawan, A. I. Lichtenstein, J. Phys.: Condens. Matter **9**, 767 (1997).

[70] A. Grüneis, M. Marsman, J. Harl, L. Schimka, and G. Kresse, G. (2009). J. of Chem. Phys. **131**, 154115 (2009).

[71] X. Ren, A. Tkatchenko, P. Rinke, and M. Scheffler, Phys. Rev. Lett. **106**, 153003 (2011).

[72] J. Paier, X. Ren., P. Rinke, G. E. Scuseria, A. Grüneis, G. Kresse, and M. Scheffler, New J. of Phys. **14**, 043002 (2012).

[73] E. Maggio, and G. Kresse, Phys. Rev. B, **93**, 235113 (2016).

[74] B. Johansson, I. A. Abrikosov, M. Aldén, A. V. Ruban, and H. L. Skriver, Phys. Rev. Lett. **74**, 2335 (1995).

[75] B. Johansson, A. V. Ruban, and I. A. Abrikosov, Phys. Rev. Lett. **102**, 189601 (2009).

[76] S. L. Dudarev, G. A. Botton, S. Y. Savrasov, C. J. Humphreys and A. P. Sutton, Phys. Rev. B **57**, 1505 (1998).

[77] R. S. Knox, *Theory of the excitons* (Academic Press, New York, 1963).

[78] J. Beille, J. Voiron, and M. Roth, Solid State Commun. **47**, 399 (1983).